\documentclass[aps,prl,showpacs,amssymb,nofootinbib,superscriptaddress,twocolumn]{revtex4-2}
\usepackage[T1]{fontenc}
\usepackage[utf8]{inputenc}
\usepackage[english]{babel}
\usepackage{bbm, bm, amsmath, dsfont, graphicx}

\usepackage{resizegather}

\begin{document}

\title{A Moment for Random Measurements}
\author{Lukas Knips}
\affiliation{Max-Planck-Institut für Quantenoptik, Hans-Kopfermann-Straße 1, 85748 Garching, Germany}
\affiliation{Department für Physik, Ludwig-Maximilians-Universität, Schellingstraße 4, 80799 München, Germany}
\affiliation{Munich Center for Quantum Science and Technology (MCQST), Schellingstraße 4, 80799 München, Germany}

\begin{abstract}
Quantum entanglement is one of the core features of quantum theory.
While it is typically revealed by measurements along carefully chosen directions, here we review different methods based on so-called \emph{random} or \emph{randomized measurements}.
Although this approach might seem inefficient at first, sampling correlations in various random directions is a powerful tool to study properties which are invariant under local-unitary transformations.
Based on random measurements, entanglement can be detected and characterized without a shared reference frame between the observers or even if local reference frames cannot be defined.

This overview article discusses different methods using random measurements to detect genuine multipartite entanglement and to distinguish SLOCC classes. 
Furthermore, it reviews how measurement directions can efficiently be obtained based on spherical designs.
\end{abstract}

\maketitle
\renewcommand{\figurename}{Figure}
\renewcommand{\tablename}{Table}


\section*{Introduction}

\emph{Entangled quantum systems are more strongly correlated than classical systems can be.}
This simplified statement proves to be remarkably rich by raising several questions, from practical ones -
``How do we measure those correlations?''
- to more subtle ones -
``What does \emph{stronger} mean here and \emph{how} much stronger are the correlations? Are there exceptions to this rule? Can we use this relation to characterize entanglement?''

While entanglement of bipartite or multipartite systems is typically detected using a set of carefully chosen measurements on all subsystems, here, we discuss a simpler and, as it turns out, yet powerful method to analyze correlations and entanglement.
There, one samples from the entire set of correlations by measuring in random directions, instead of considering a specific set of correlations.
This type of measurement is often called a \emph{random} or \emph{randomized measurement}.
Depending on the context, these can be either controlled measurements without a shared reference frame, measurements without active control over (but knowledge of) the measurement direction, or even measurements with neither control nor knowledge about the measurement direction.

Previous works showed the violation of Bell-type inequalities without the need for a shared reference frame~\cite{liang_nonclassical_2010,palsson_experimentally_2012,shadbolt_generating_2012,shadbolt_guaranteed_2012,de_rosier_multipartite_2017,fonseca_survey_2018}, but with the ability to repeat previously conducted measurements.
Random measurements in the sense of measurements without control have been used in the context of many-body systems~\cite{elben_renyi_2018,elben_statistical_2019,brydges_probing_2019,elben_many-body_2020,elben_mixed-state_2020}, for the verification of quantum devices~\cite{elben_cross-platform_2020}, for the detection of entanglement~\cite{tran_quantum_2015,tran_correlations_2016,dimic_single-copy_2018,saggio_experimental_2019}, for the prediction of fidelities, entanglement entropies and various other properties~\cite{huang_predicting_2020} as well as for the characterization and classification of genuine multipartite entanglement~\cite{saggio_experimental_2019,ketterer_characterizing_2019,ketterer_entanglement_2020,knips_multipartite_2020}.
Recently, it was shown that even bound entangled states, i.e., states so weakly entangled that their entanglement is not recognized by the PPT (positive partial transpose) criterion~\cite{peres_separability_1996,horodecki_separability_1996}, can be characterized in a reference-frame independent manner~\cite{imai_multiparticle_2020}.

In this perspective article, we will discuss a work of Ketterer et al., which recently appeared in \textit{Quantum}~\cite{ketterer_entanglement_2020}.
Before we will review their means of entanglement detection and classification, we will introduce the general concept of random measurements, provide an intuitive understanding for them and give context by discussing other methods for detection and classification of entanglement in this scenario.
Finally, we will discuss their approach for selecting local measurement directions based on spherical $t$-designs.

\section*{Scenario}

\begin{figure}[h!]
\centering
\includegraphics[width=0.45\textwidth]{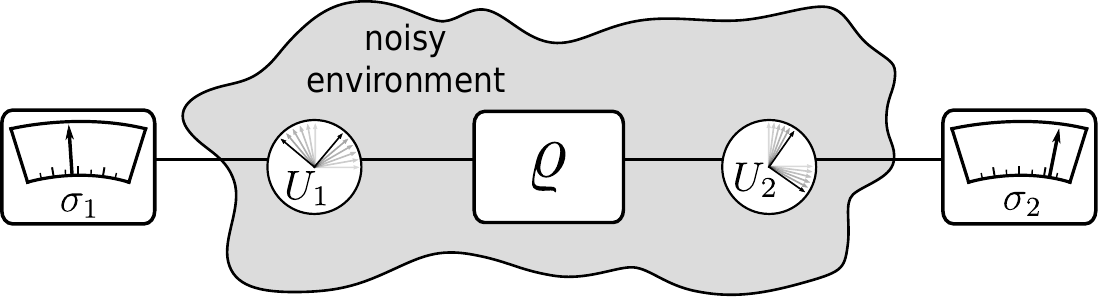}
\caption{Concept of random measurements: a quantum state is distributed and is assumed to undergo (unknown) local unitary transformations.}
\label{fig:concept}
\end{figure}

To illustrate the scenario, we can first think of a two-qubit state whose subsystems are sent to two observers via unitary but unknown quantum channels as shown in Fig.~\ref{fig:concept}. Hence, the goal is to characterize the quantum state as well as possible despite the lack of a shared reference frame.
Interestingly, although the correlation value of the outcomes of both observers (which quantifies the probability for both results being equal or opposite) is random as it depends on the respective measurement directions, the \emph{distribution} of correlation values turns out to be a useful resource for describing the state.

If the unknown unitary transformations of the quantum channels are furthermore time-dependent, i.e., the channels are \emph{fluctuating} or \emph{noisy}, one could naively perform measurements in any fixed direction.
However, the type of fluctuations strongly influences the measurement results: are the fluctuations distributed uniformly (according to the Haar measure) or are we oversampling some and undersampling other directions?
To mitigate this problem and avoid any dependence on the type of noise of the quantum channels, the measurement directions can themselves be chosen Haar-randomly.
In this way, the concatenation of the quantum channel of the noisy environment featuring possibly biased noise with the channel corresponding to a Haar random distribution of unitary transformations due to intentional rotation removes any bias.

\section*{Intuitive Picture}

\begin{figure}[h!]
\centering
\includegraphics[width=0.45\textwidth]{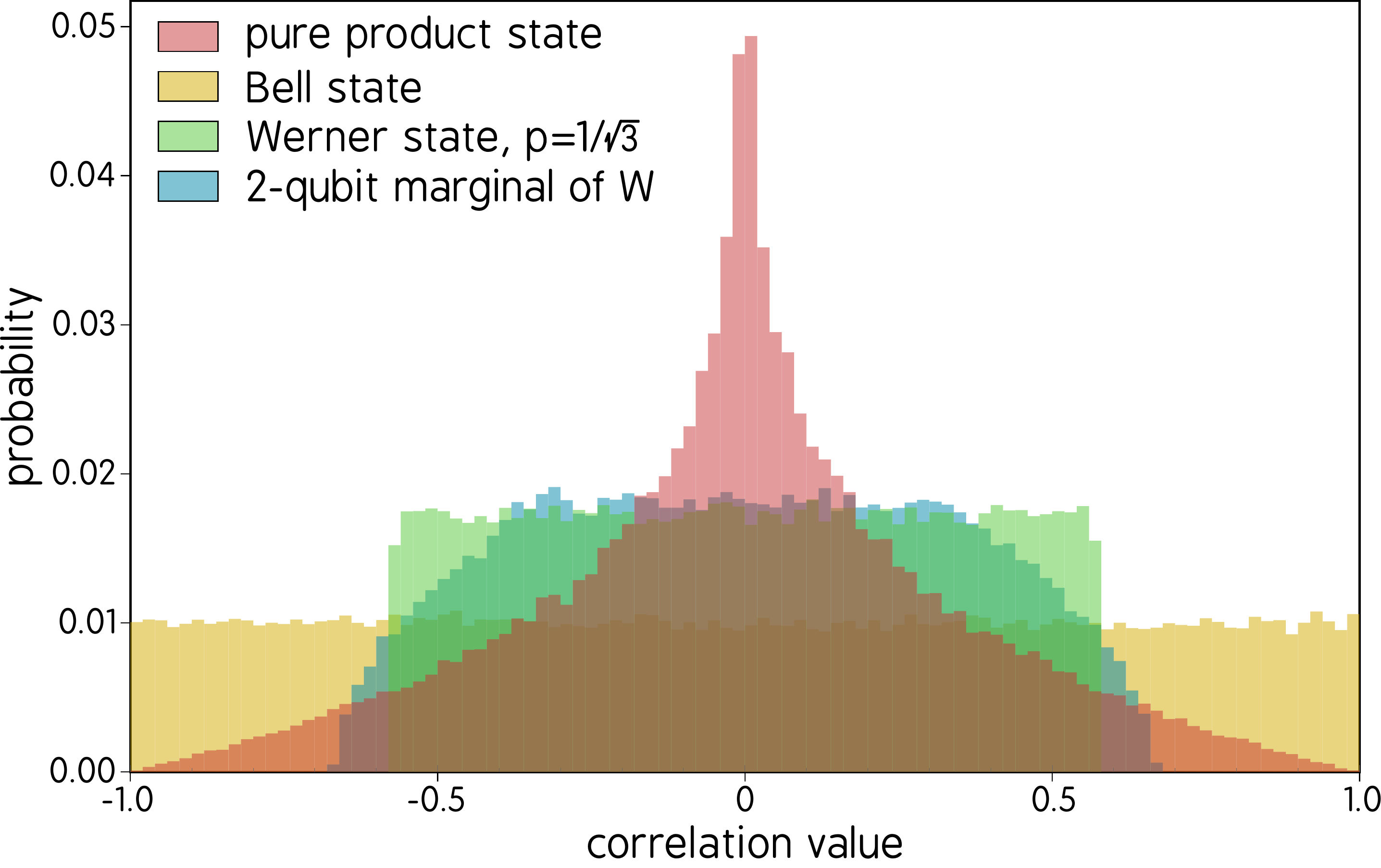}
\caption{Measuring different quantum states in many different randomly chosen directions may lead to distinct distributions of correlations.}
\label{fig:distributions}
\end{figure}

To establish an intuitive understanding how entangled and separable quantum states behave in this scenario and how we can make use of the stronger correlations of an entangled state, let us do a short numerical experiment.
Take the pure product state $|\psi^{\mathrm{prod}}\rangle=|00\rangle$ and a Bell state, e.g., $|\psi^-\rangle=(|01\rangle-|10\rangle)/\sqrt{2}$.
Now repeatedly draw two $2\times2$ unitary matrices according to the Haar measure (see, e.g., \cite{mezzadri_how_2007} for a practical recipe).
Apply the first (second) of those unitaries to the first (second) qubit of both states and evaluate, e.g., $\langle\sigma_z\otimes\sigma_z\rangle$, i.e., the correlation value in the $zz$-direction.
A histogram of those two distributions shows the very distinct behavior of the product and the maximally entangled state, as shown in figure \ref{fig:distributions}.
The maximally entangled state results in a uniform distribution of the expectation values of the correlations, whereas small absolute values are much more probable than large ones for the product state.

\begin{figure}[h!]
\centering
\includegraphics[width=0.45\textwidth]{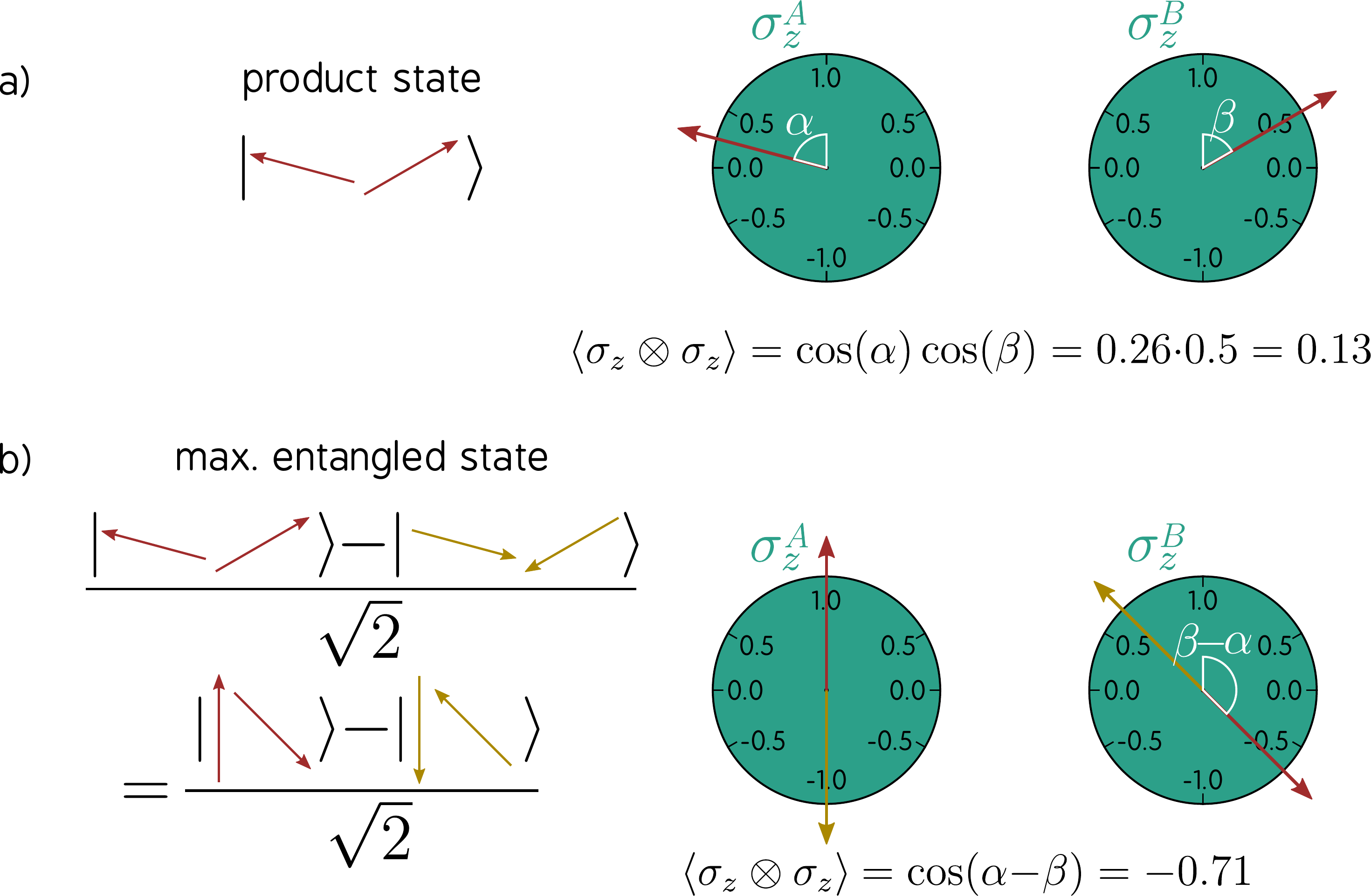}
\caption{If we compare a pure two-qubit product state with a maximally entangled state, we find that correlation values for the former state depend on two angles, whereas only a single angle suffices to describe the correlations of the latter state.}
\label{fig:arrows}
\end{figure}

For those two states, it is still easy to understand the origin of the corresponding distributions.
Consider the schematic arrow diagrams in Fig. \ref{fig:arrows}, where in a) the two red arrows represent the spins of a product state after applying arbitrary local unitary (LU) transformations each parametrized here by a single angle.
A measurement of $\sigma_z\otimes\sigma_z$ is just given by the product of the results, which are obtained by the projection onto the $\sigma_z$ directions.
Both angles, $\alpha$ and $\beta$, have to be close to $0$ or to $\pi$ to give a large absolute correlation value.
In the example, we show the case of $\alpha=75^\circ$ and $\beta=60^\circ$ leading to a $zz$-correlation of about $0.13$.

For the maximally entangled state in Fig.~\ref{fig:arrows} b), the situation is very different.
We illustrate the state as a superposition of both red arrows with both yellow arrows.
By applying an LU transformation of the form $U\otimes U$ on both qubits such that, say, the first qubit is aligned with its measurement direction (i.e., such that $U$ is a rotation by an angle $-\alpha$), the state does not change (up to a global phase factor).
The expectation value of $\sigma_z\otimes\sigma_z$ therefore now obviously only depends on the single relative angle $\beta-\alpha$ instead of the two angles $\alpha$ and $\beta$.
For the same angles as in the product-state case, we now obtain a $zz$-correlation of about $-0.71$.

Formally, we can obtain the distributions of correlations $E$ by integration over the Bloch spheres $S^{2}\times S^{2}$ as
\begin{align}
&p_{\mathrm{prod}}(E) = \frac{1}{(4\pi)^2} \int_{S^2} {\mathrm{d}U_1}\int_{S^2} {\mathrm{d}U_2} \delta(E-E_{\mathrm{prod}}(U_1,U_2)) \nonumber \\
&= \frac{1}{4} \int_0^{\pi} \sin(\theta_1) {\mathrm{d}\theta_1}\int_0^{\pi} \sin(\theta_2){\mathrm{d}\theta_2} \delta(E - \cos(\theta_1) \cos(\theta_2)  ) \nonumber \\
&= -\frac{1}{2} \log(|E|),\\
&p_{\mathrm{Bell}}(E) = \frac{1}{(4\pi)^2} \int_{S^2} {\mathrm{d}U_1}\int_{S^2} {\mathrm{d}U_2} \delta(E-E_{\mathrm{Bell}}(U_1,U_2)) \nonumber \\
&= \frac{1}{4} \int_0^{\pi} \sin(\theta_1) {\mathrm{d}\theta_1}\int_0^{\pi} \sin(\theta_2){\mathrm{d}\theta_2} \delta(E - \cos(\theta_1-\theta_2)  ) \nonumber \\
&= \frac{1}{2}.
\end{align}
In the histograms of Fig.~\ref{fig:distributions}, we have also shown the distributions of two other states.
The Werner state, as the mixture of the Bell state (corresponding to a uniform distribution) and the maximally mixed state (corresponding to a Dirac delta peak at $0$ as the maximally mixed state $\mathbbm{1}/4$ always results in $\langle \sigma_z\otimes\sigma_z\rangle=0$, independently of the measurement directions) with mixing parameter $p$, results in a uniform distribution in the range $[-p,p]$ (shown above for $p=1/\sqrt{3}$), i.e., mixing white noise to a Bell state bounds the possible correlations in this sense.
The two-qubit marginal of a tripartite $W$ state, however, gives yet another distinct distribution.

Due to the nature of the measurements, two quantum states which are equivalent up to local unitary transformations will show the same distribution of correlations.
Therefore, all pure two-qubit product states result in a logarithmic distribution, while, for example, all maximally entangled two-qubit states result in a uniform distribution.

\section*{Statistical Moments}

To characterize such distributions of correlations, statistical moments have proven to be powerful.
The $t$-th moment of a probability distribution $p(x)$ is given by
\begin{equation}
m^{(t)} = \int x^t p(x) {\mathrm{d}x}.
\end{equation}
The first moment ($t=1$) is the mean value.
The next lower
\emph{centralized}
moments (i.e., the moments after shifting the distribution around its mean) are the variance, the skewness and the kurtosis.
As we are dealing with a symmetric distribution here, the mean value vanishes and, hence, the centralized moments are identical to the moments themselves.
In our scenario, where all moments are finite and the moment-generating function has a positive radius of convergence, knowing all moments allows one to uniquely determine the distribution~\cite{papoulis_probability_1984}.
Random measurements can be used to obtain the statistical moments of the distribution of expectation values.
The scheme naturally generalizes to the case of more than two parties~\cite{van_enk_measuring_2012,klockl_characterizing_2015,tran_quantum_2015} and is not limited to qubits \cite{tran_quantum_2015,klockl_characterizing_2015,tran_correlations_2016}.

By considering the measurement results on a subset of parties, also distributions of correlations of marginal states can be retrieved.
As the second moment is the first non-trivial one here, we will from now on denote it just as $m:= m^{(2)}$ and specify the respective set of parties it pertains to using a subscript.
The combined information of the second moments of the
\emph{full}
distribution ($m_{1,2,\dots,n}$) involving all $n$ observers together with those of all marginal states ($m_{1}$, $m_{1,2}$, $m_{1,2,3}$, $\dots$, $m_{(n-1),n}$, $\dots$, $m_{n}$) allows us to determine the purity of an $n$-qubit state~\cite{van_enk_measuring_2012,lawson_reliable_2014,klockl_characterizing_2015,knips_multipartite_2020} as
\begin{align}
\operatorname{tr}{\varrho^2}=\frac{1}{2^n} \sum_{\mathcal{A} \in \mathbbm{P}(\mathcal{S})}{ 3^{|\mathcal{A}|} \, m_\mathcal{A}},
\end{align}
where $\mathbbm{P}(\mathcal{S})$ is the set of all subsets of $\mathcal{S}=\{1,\ldots,n\}$ and $|\mathcal{A}|$ denotes the cardinality (number of elements) of the set $\mathcal{A}$.
Here, $m_\mathcal{A}$ is the second moment of the distribution involving the observers given by $\mathcal{A}$.
In other words, the purity is given by the weighted sum of second moments of the full distribution and all marginal distributions.
In the remainder of this perspective, we will discuss how to detect and to characterize entanglement based on statistical moments.

\section*{Detecting Entanglement}

A remarkably simple method to detect entanglement was presented in~\cite{tran_quantum_2015}.
For any pure, fully separable $n$-qubit state $|\psi\rangle=|\psi_1\rangle\otimes|\psi_2\rangle\otimes\dots|\psi_n\rangle$, the length of correlations (sum of the correlations of all basis directions squared) is $1$.
If
\emph{each}
observer $j$ aligns their measurement apparatus along $|\psi_j\rangle$, they jointly observe a correlation of $1$, whereas any orthogonal measurement direction will result in the loss of correlated outcomes.
This relation holds independently of the number of parties and can be adapted for arbitrary dimensions.
Choosing $M$ measurement directions completely randomly with $K$ repetitions each allows to estimate the length of correlations.
If one significantly overcomes the threshold of a product state, one can, after proper statistical analysis, conclude that the state must carry \emph{some} entanglement.

\begin{figure*}[!ht]
\centering
\includegraphics[width=0.95\textwidth]{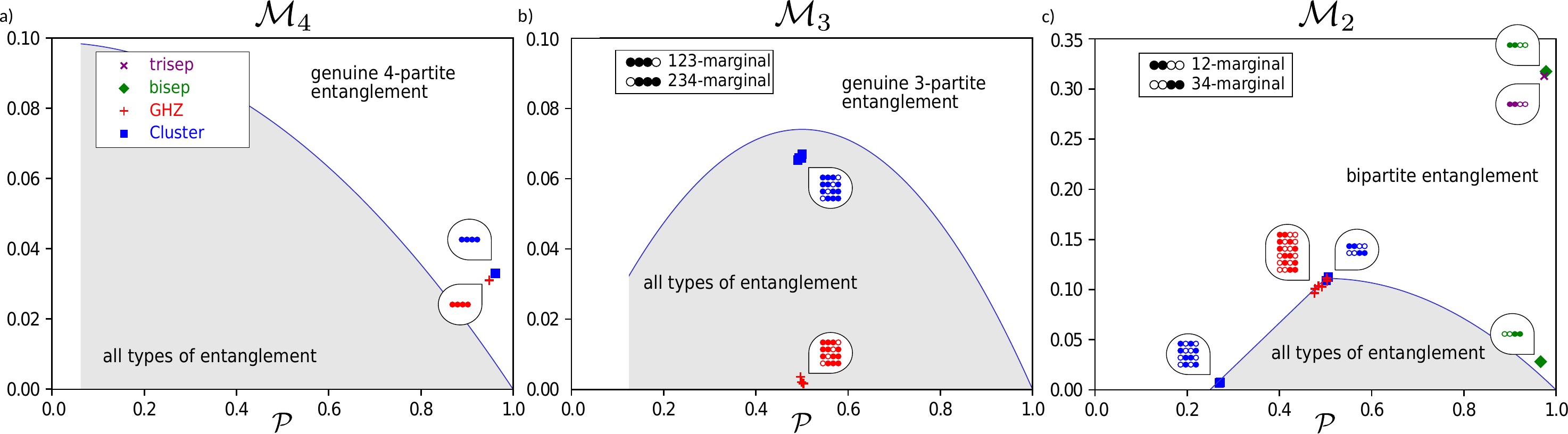}
\caption{For different four-qubit states, it can be evaluated how much the distribution of correlations can be described by the product of moments of its subsystems.
%
%
If one finds the quantity ${\mathcal M}_{4}$ above its purity-dependent bound, the state is proven to be genuinely fourpartite entangled as shown for the four-qubit GHZ and the four-qubit linear Cluster state.
For the tri- and biseparable states, no genuine fourpartite entanglement is detected.
However, the marginals of the first two qubits (for both states) as well as of the last two qubits (for the biseparable state) give values of ${\mathcal M}_{2}$ above the respective bound, indicating entanglement within those marginals.}
\label{fig:gme}
\end{figure*}

Subsequent work~\cite{tran_correlations_2016} studied the length of correlations of various genuinely multipartite entangled states and extended the results of~\cite{tran_quantum_2015} to mixed states.
There, entanglement detection based on a single measurement setting is derived explicitly.
However, it was found that the length of correlations (alone) is not an entanglement measure as it can increase under local operations and classical communication (LOCC).
Nevertheless, random measurements can be used to witness genuine multipartite entanglement, i.e., entanglement truly involving all parties, as we will see below.
Moreover, classes of states inequivalent under stochastic local operations and classical communication (SLOCC)~\cite{dur_three_2000,acin_classification_2001} can be distinguished in this way, as we will discuss.

Yet another approach for entanglement detection is used in \cite{dimic_single-copy_2018,saggio_experimental_2019}.
There, a measurement direction is randomly drawn from a specific set.
In addition, a framework for the probabilistic use of entanglement witnesses is provided.
With this, entanglement can in some cases be detected with a single copy with high confidence~\cite{dimic_single-copy_2018} or different classes of entanglement can be discriminated using a few copies of the state in a more general approach~\cite{saggio_experimental_2019}.

Also note that it is common that entanglement criteria are sufficient, but not necessary.
For example, any method, whether based on random measurements or on fully controlled ones, that is only using \emph{full} correlations, i.e., correlations between all $n$ parties without inspecting correlations between subsets of parties, will miss some entangled states: There are genuinely $n$-partite entangled states without any correlations between all $n$ parties~\cite{schwemmer_genuine_2015,tran_genuine_2017,klobus_higher_2019}, which therefore will result in a Dirac-delta peak for the full distribution and are on that level indistinguishable from white noise.

\section*{Detecting Genuine Multipartite Entanglement}

\subsection*{Using Second Moments of Marginals}
In \cite{knips_multipartite_2020}, the second moments of the distributions are used for the detection of genuine multipartite entanglement (entanglement which truly involves all parties and cannot be broken down into pure biseparable states or even mixtures of states biseparable with respect to different bipartitions).
The key ingredient of that strategy is to relate the second moment of the distribution of correlations to the second moments of marginal distributions and test that quantifier with a purity-dependent threshold.
For a pure product state $|\psi_{\mathcal S}\rangle=|\psi_{{\mathcal A}_1}\rangle\otimes|\psi_{{\mathcal A}_2}\rangle$, the second moment of the distribution of correlations when considering the set of parties $\mathcal S$ factorizes into the second moments of the corresponding marginal distributions, i.e., $m_{{\mathcal S}}=m_{{\mathcal A}_1}m_{{\mathcal A}_2}$.
For a pure state, genuine multipartite entanglement can therefore be detected by verifying $m_{{\mathcal S}}>m_{{\mathcal A}_1}m_{{\mathcal A}_2}$ for every possible bipartition ${\mathcal A}_1|{\mathcal A}_2$ such that ${\mathcal S}={\mathcal A}_1\cup{\mathcal A}_2$.

Unfortunately, for non-pure biseparable states, the second moment of the full distribution might be larger than the product of the marginals' second moments.
To mitigate this problem, purity-dependent bounds for detecting genuine multipartite entanglement can be used.
For an $n$-qubit state we can consider
\begin{equation}
{\mathcal M}_{n} := m_{{\mathcal S}} - \frac{1}{2} \sum_{{\mathcal A} \in \{\mathbbm{P}({\mathcal S}) \setminus ({\mathcal S} \cup \varnothing)\}}{ m_{\mathcal A} m_{{\mathcal S} \setminus {\mathcal A}}}
\end{equation}
to capture by how much the full distribution's second moment can be expressed in terms of the marginals' ones.
For example, it has been found~\cite{knips_multipartite_2020} that in the case of $n=4$ qubits, all biseparable states fulfill the relation
\begin{eqnarray}
{\mathcal M}_{4}
\le \tfrac{8}{81}(1-\operatorname{tr}\varrho^2).
\label{eq:M4}
\end{eqnarray}
A violation of the latter inequality therefore indicates genuine fourpartite entanglement.

In Fig.~\ref{fig:gme}, the values of ${\mathcal M}_{4}$ based on the distribution of fourpartite correlations for experimental photonic states (four qubits encoded in polarization and path degrees of freedom of two photons from type-I spontaneous parametric down-conversion, see~\cite{knips_multipartite_2016} for details on the setup) are compared with the purity-dependent threshold given by Eq.~(\ref{eq:M4})~\cite{knips_multipartite_2020}.
As the red (four-qubit GHZ state) and blue (linear four-qubit Cluster state) data points are above the threshold, those states are shown to be genuinely fourpartite entangled.
The distributions of a prepared triseparable state [$\propto \left(|00\rangle+|11\rangle\right)\otimes|0\rangle\otimes|0\rangle$] and of a biseparable state [$\propto \left(|00\rangle+|11\rangle\right)\otimes\left(\sin\varphi\,|00\rangle+\cos\varphi\,|11\rangle\right)$ with $\varphi\approx0.2$], however, contain only one and two bipartite entangled marginals, respectively, as one finds by considering ${\mathcal M}_{3}$ for all tripartite and ${\mathcal M}_{2}$ for all bipartite marginals of those states.
Thus, this approach not only allows to detect genuine multipartite entanglement for mixed states, but generally provides insights into the entanglement structure.

\subsection*{Using Higher Order Moments}
The combination of moments of various orders may in general capture more information about a distribution of correlations and, hence, about the underlying quantum state than restricting the analysis to moments of second order, only.
In two recent works~\cite{ketterer_characterizing_2019,ketterer_entanglement_2020}, the latest of which was published in \emph{Quantum}~\cite{ketterer_entanglement_2020}, Ketterer et al. use a combination of the second and the fourth moment, denoted there by $\mathcal{R}^{(2)}$ and $\mathcal{R}^{(4)}$ (with $\mathcal{R}^{(t)}\equiv m^{(t)}$), respectively.
Obviously, those moments are not entirely independent of each other.
For example, a vanishing second moment $\mathcal{R}^{(2)}$ indicates a Dirac delta distribution, which in turn requires $\mathcal{R}^{(4)}$ to vanish.
They discuss possible combinations of those two moments for two, three and four qubits.
In a combined analytical and numerical study, the authors identify regions in an $\mathcal{R}^{(2)}$-$\mathcal{R}^{(4)}$ plane which allow them to directly indicate that a state is, e.g., biseparable or that it cannot belong to a specific SLOCC class.

\begin{figure}[h!]
\centering
\includegraphics[width=0.45\textwidth]{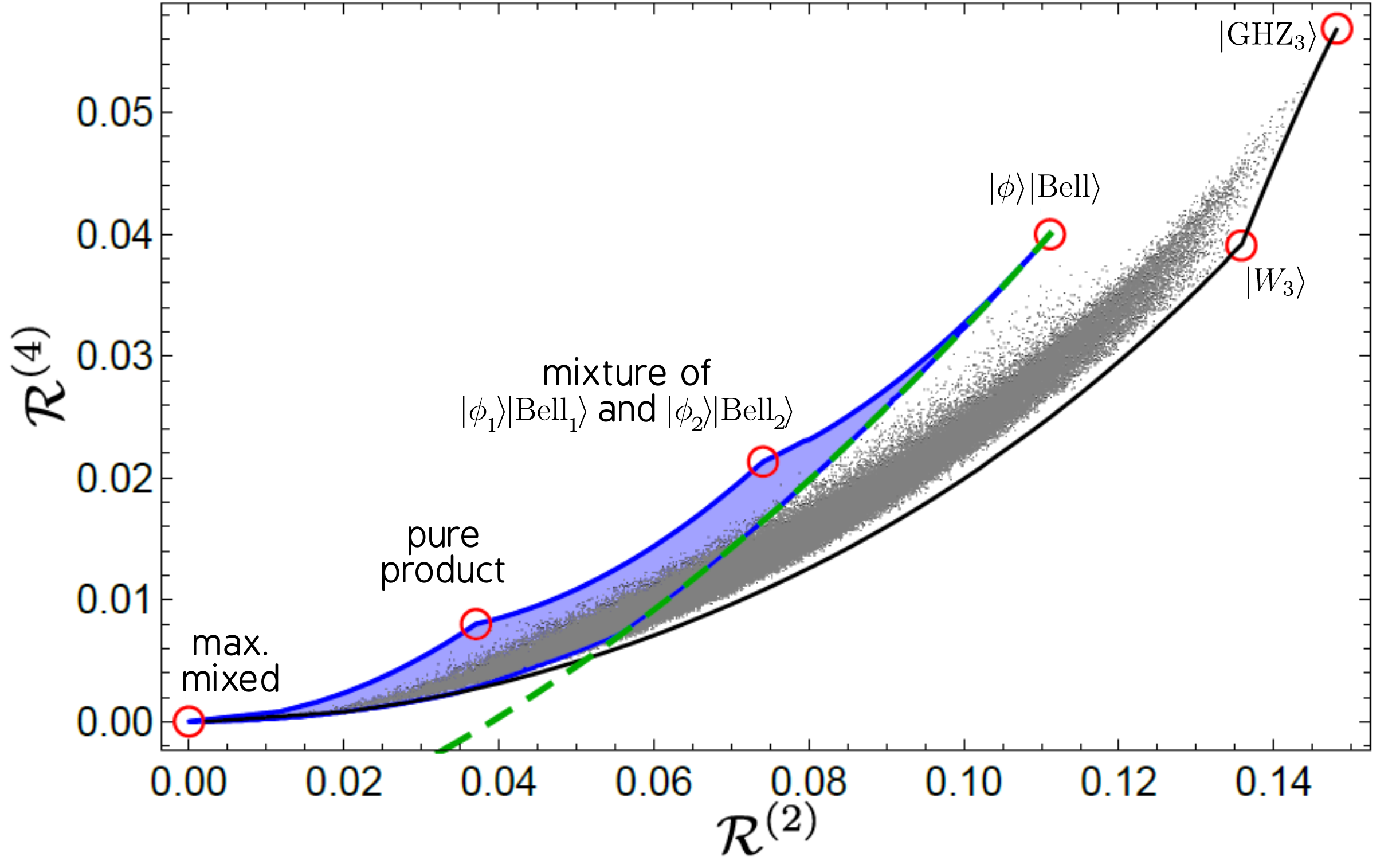}
\caption{Three-qubit states are being analyzed based on their second and fourth moments $\mathcal{R}^{(2)}$ and $\mathcal{R}^{(4)}$, respectively. The blue-shaded region marks biseparable states. Therefore, the green-dashed line as described by Eq.~(\ref{eq:threequbit_bisep}) can be used to detect genuine tripartite entanglement. Figure from~\cite{ketterer_entanglement_2020}.}
\label{fig:3qubitgme}
\end{figure}

In Fig.~\ref{fig:3qubitgme}, three-qubit states are sampled and represented in the $\mathcal{R}^{(2)}$-$\mathcal{R}^{(4)}$ plane.
The blue-shaded area is outlined by different LU-inequivalent types of biseparable states.
Ketterer et al. propose the inequality
\begin{equation}
\mathcal{R}^{(4)} \geq \frac{1}{425}\left[972 \left(\mathcal{R}^{(2)}\right)^2+90\mathcal{R}^{(2)}-5\right],\label{eq:threequbit_bisep}
\end{equation}
which is shown as the green dashed line, as demarcation between biseparable and genuine multipartite entangled states.
Therefore, states for which the fourth moment $\mathcal{R}^{(4)}$ is below this threshold are not biseparable and are hence shown to be genuinely multipartite entangled.

\section*{Witnessing SLOCC classes}
\begin{figure}[h!]
\centering
\includegraphics[width=0.45\textwidth]{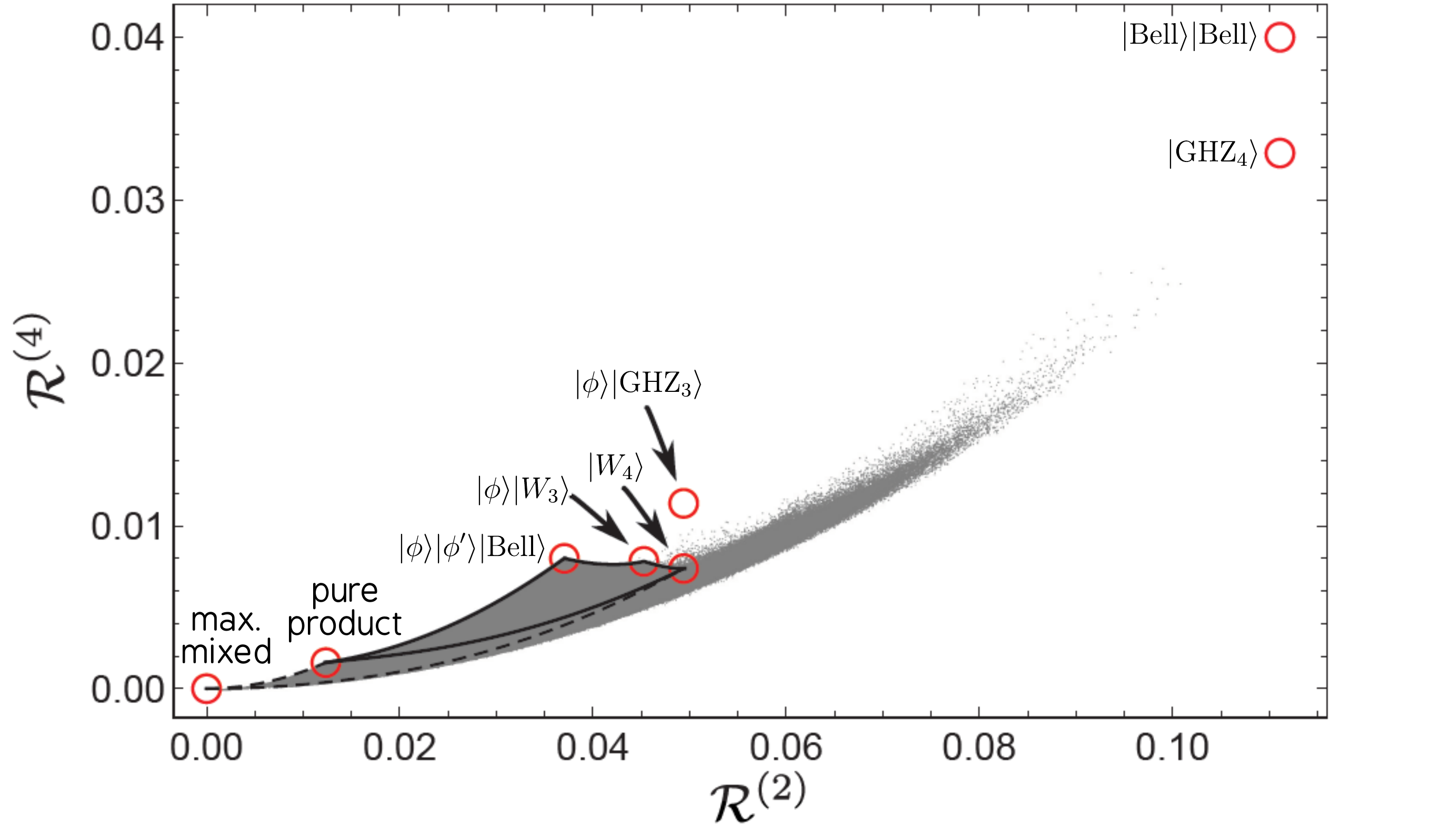}
\caption{The second moment $\mathcal{R}^{(2)}$ and the fourth moment $\mathcal{R}^{(4)}$ allow to discriminate four-qubit states as shown here. With this, also different entanglement classes can be distinguished. The solid black line encloses the SLOCC class of $W$ states, whereas the dashed line surrounds the convex hull of the $W$ states. Figure from~\cite{ketterer_entanglement_2020}.}
\label{fig:slocc}
\end{figure}

Furthermore, in \cite{ketterer_entanglement_2020} the authors discuss moments of distributions of correlations for witnessing SLOCC classes, which allows to decide if a state is reversibly convertible into, say, a $W$ state.
Figure \ref{fig:slocc} shows sampled four-qubit states in the $\mathcal{R}^{(2)}$-$\mathcal{R}^{(4)}$ plane.
The region with the solid border contains states of the $\mathcal{W}^{(4)}$ class, i.e., the SLOCC class of $W$ states, whereas the region surrounded by the dashed line encompasses its convex hull $\operatorname{Conv}(\mathcal{W}^{(4)})$.
States whose moments lie outside of the regions enclosed by the solid and dashed lines are shown not to belong to the SLOCC classes $\mathcal{W}^{(4)}$ and $\operatorname{Conv}(\mathcal{W}^{(4)})$, respectively.

We already see from Fig.~\ref{fig:slocc} that $\mathcal{R}^{(2)}$ might be very helpful for witnessing that a state is not a member of the mixed $W$ class $\operatorname{Conv}(\mathcal{W}^{(4)})$.
In contrast, additional consideration of $\mathcal{R}^{(4)}$ does not improve the ability for detection significantly.
More generally, the authors of~\cite{ketterer_entanglement_2020} derive a witness for $\operatorname{Conv}(\mathcal{W}^{(n)})$ for an arbitrary number of $n$ qubits.
If the second moment $\mathcal{R}^{(2)}$ of a distribution of correlations is larger than
\begin{equation}
\chi^{(n)}:=\frac{5-\frac{4}{n}}{3^n},
\end{equation}
the $n$-qubit state is not a member of the mixed $W$ class $\operatorname{Conv}(\mathcal{W}^{(n)})$~\cite{ketterer_entanglement_2020}.
For $4$ qubits, the threshold is $\chi^{(4)}=4/81\approx0.049$.
Hence, all states with $\mathcal{R}^{(2)}>\chi^{(4)}=4/81$, i.e., on the right-hand side of $|\mathrm{W}_4\rangle$ and $|\phi\rangle\otimes|\mathrm{GHZ}_3\rangle$ in Fig.~\ref{fig:slocc}, are shown not to belong to the mixed $W$ class.

Please note that this is not a statement about genuine multipartite entanglement.
For example, both the biseparable state $|\mathrm{Bell}\rangle\otimes|\mathrm{Bell}\rangle$ and the genuinely fourpartite entangled state $|\mathrm{GHZ}_4\rangle$ are outside of this region as both are not members of $\operatorname{Conv}(\mathcal{W}^{(4)})$.

\section*{Spherical Designs}
Ketterer et al.~\cite{ketterer_characterizing_2019,ketterer_entanglement_2020} not only use higher-order moments to detect and characterize entanglement, but they also follow a different approach for selecting measurement directions.
Up to this point of the discussion it has been assumed that the distributions of correlations are obtained by sampling over a large set of random directions, where the distribution of directions should follow a Haar random distribution.
Whereas in, e.g., \cite{knips_multipartite_2020} the sampling was done over a large set of Haar randomly distributed measurement directions to describe the distributions of correlations, the approach of Ketterer et al. allows one to fix the number of measurement directions if a specific moment is to be calculated.
This significantly reduces the measurement effort at the cost of requiring active control over the local measurement directions.

Using the notation of Ref.~\cite{ketterer_entanglement_2020}, the $t$-th moment $\mathcal{R}^{(t)}$ of the distribution of correlations of an $n$-qubit state can be obtained from
\begin{align}
\mathcal{R}^{(t)} &= \!\!\int\limits_{\mathcal{U}(2)} \!\!\!\mathrm{d}\eta (U_1) \cdots\!\! \int\limits_{\mathcal{U}(2)}\!\!\! \mathrm{d}\eta(U_n)\langle U_1\sigma_z U_1^\dagger\otimes \dots \otimes U_n\sigma_z U_n^\dagger \rangle^t \nonumber \\
&= \frac{1}{\left(4\pi\right)^n} \int_{S^2} \mathrm{d}\vec{u}_1 \cdots \int_{S^2} \mathrm{d}\vec{u}_n E(\vec{u}_1,\dots,\vec{u}_n)^{t}, \label{eq:tth_moment_integration}
\end{align}
where $E(\vec{u}_1,\dots,\vec{u}_n):=\langle \sigma_{\vec{u}_1}\otimes\dots\otimes\sigma_{\vec{u}_n} \rangle$ denotes the correlation along specific local measurement directions $\vec{u}_i$ with $\sigma_{\vec{u}_i}=\vec{u}_i\cdot\vec{\sigma}$, where $\vec{\sigma}=\left(\sigma_x,\sigma_y,\sigma_z\right)^{T}$ is a vector of the Pauli matrices $\sigma_x$, $\sigma_y$ and $\sigma_z$, while $\eta$ and $\mathrm{d}\vec{u}_i=\sin\theta_i\mathrm{d}\theta_i\mathrm{d}\varphi_i$ are the Haar measure on the unitary group $\mathcal{U}(2)$ and the uniform measure on the Bloch sphere $S^2$, respectively.

To determine the average of a homogeneous polynomial $P_{t^\prime}:S^2\rightarrow\mathbb{R}$ of order $t^\prime$ over the Bloch sphere $S^2$, it is sufficient to sample a finite set of points as shown in~\cite{ketterer_characterizing_2019,ketterer_entanglement_2020}.
For that, they use a so-called spherical $t$-design in dimension three which is defined by the finite set of points $\{\vec{u}_i|i=1,\dots,L^{(t)}\}\subset S^2$ such that
\begin{equation}
\int_{S^2} \mathrm{d}\vec{u} P_{t^\prime}(\vec{u}) = \frac{1}{L^{(t)}}\sum_{k=1}^{L^{(t)}}P_{t^\prime}(\vec{e}_k)
\end{equation}
holds for all homogeneous polynomials of order $t^\prime$ with $t^\prime\leq t$.
Hence, for the respective spherical $t$-design, $L^{(t)}$ determines the number of measurement directions to consider.
Using this framework, Ketterer et al. evaluate the $t$-th moment of the correlations of an $n$-qubit state as
\begin{equation}
\mathcal{R}^{(t)} = \frac{1}{\left(L^{(t)}\right)^n}\sum_{k_1,\dots,k_n=1}^{L^{(t)}} \langle \sigma_{\vec{u}_1}\otimes\dots\otimes\sigma_{\vec{u}_n} \rangle^{t},
\end{equation}
instead of using the integration as in Eq.~(\ref{eq:tth_moment_integration}).
Although they also show a similar derivation for qudit states employing
\emph{unitary}
$t$-designs, we here restrict our discussion to the qubit case using
\emph{spherical}
$t$-designs.

\begin{figure}[h!]
\centering
\includegraphics[width=0.45\textwidth]{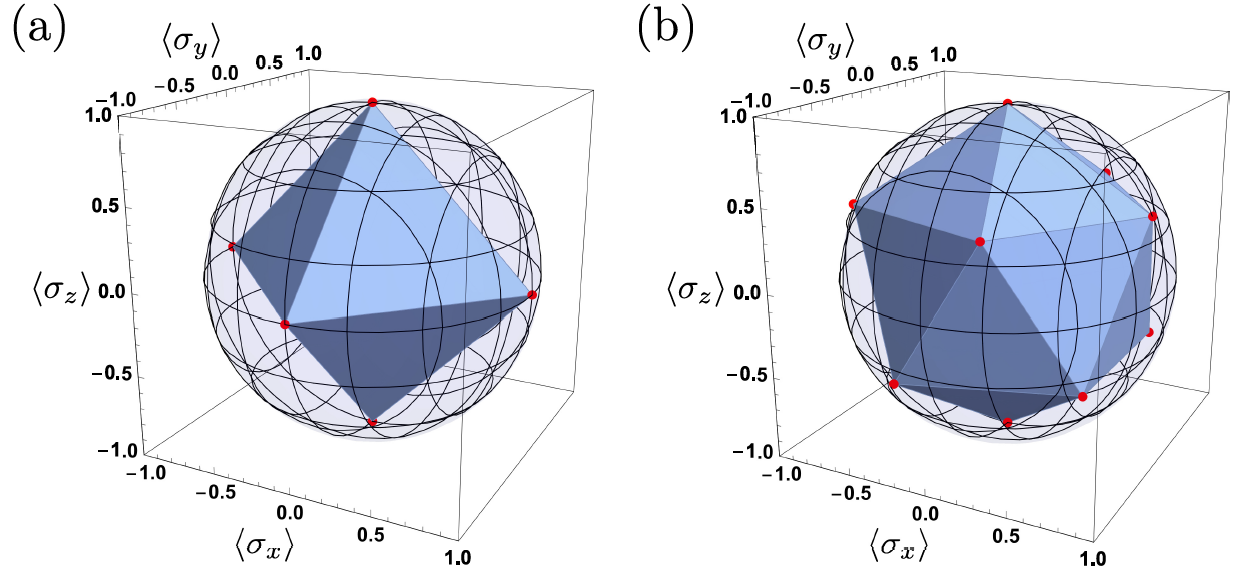}
\caption{Spherical $t$-designs as the shown $3$-design (a) and $5$-design (b) can be used to find measurement directions when calculating moments of a specific order. Figure from~\cite{ketterer_entanglement_2020}.}
\label{fig:sphericaldesign}
\end{figure}

In Fig.~\ref{fig:sphericaldesign}, the $L^{(3)}=6$ directions $\{\pm \vec{e}_i|i=x,y,z\}$ of a spherical $3$-design as well as the $L^{(5)}=12$ directions of a $5$-design are shown.
If one is only interested in the second moment (polynomial of order $2$), the $6$ local measurement directions of $L^{(3)}$ are sufficient.
Moreover, as even-order moments are invariant under a parity transformation of the measurement direction, skipping $\{-\vec{e}_i|i=x,y,z\}$ does no harm.
For obtaining the fourth moment, $L^{(5)}/2=6$ local measurement directions are suitable.
With this method, the selection of measurement directions from the pseudo-random process of a spherical design allows one to mimic uniform averages over the sphere~\cite{ketterer_entanglement_2020}.

\section*{Conclusion and Future Research}
In this perspective, we have discussed what we can learn from random measurements despite the lack of shared or even local reference frames.
Distributions of correlations can reveal entanglement and exclude any type of separability.
Although considering only the second moment of these distributions is not sufficient for constructing an entanglement measure, it can be used for witnessing SLOCC classes as recently shown in \cite{ketterer_characterizing_2019,ketterer_entanglement_2020}.
As random measurements are inherently not sensitive to local unitary transformations, they do not divert one's gaze from LU-invariant properties.

Random measurements turn out to be a powerful tool for entanglement detection and classification.
Here, we did not elaborate on statistical errors involved in those measurements, which require some further research.
Also, we focused on quantum
\emph{states}.
Of course, random measurements are also of interest for characterizing quantum
\emph{processes}.
Another open question is the tomographic reconstruction using random measurements: Which states can be discriminated and what information will stay hidden?
Also, it is worth to discuss how and to what degree random measurements can be employed for applications such as quantum metrology.

\section*{Acknowledgments}
I am grateful to Tomasz Paterek, Jasmin D. A. Meinecke, Karen Wintersperger, and Nicolai Friis for their helpful comments and suggestions for improvements of the manuscript.

\end{document}